\documentclass[10pt,a4paper]{article}
\usepackage[top=2cm,left=7.6cm,footskip=1cm]{geometry}

\usepackage{amsmath,amssymb}

\usepackage{changepage}

\usepackage[utf8x]{inputenc}

\usepackage{textcomp,marvosym}

\usepackage{cite}

\usepackage{nameref,hyperref}

\usepackage[right]{lineno}

\usepackage{microtype}
\DisableLigatures[f]{encoding = *, family = * }

\usepackage[table]{xcolor}

\usepackage{array}

\newcolumntype{+}{!{\vrule width 2pt}}

\newlength\savedwidth

\raggedright
\usepackage[aboveskip=1pt,labelfont=bf,labelsep=period,justification=raggedright,singlelinecheck=off]{caption}

\makeatletter
\renewcommand{\@biblabel}[1]{\quad#1.}
\makeatother

\usepackage{lastpage,fancyhdr,graphicx}
\usepackage{epstopdf}
\usepackage{multirow}
\usepackage{subcaption}

\usepackage{listings}
\usepackage{color}
\definecolor{lightgray}{rgb}{0.95, 0.95, 0.95}
\definecolor{darkgray}{rgb}{0.4, 0.4, 0.4}
\definecolor{purple}{rgb}{0.65, 0.12, 0.82}
\definecolor{editorGray}{rgb}{0.95, 0.95, 0.95}
\definecolor{editorOcher}{rgb}{1, 0.5, 0} % #FF7F00 -> rgb(239, 169, 0)
\definecolor{editorGreen}{rgb}{0, 0.5, 0} % #007C00 -> rgb(0, 124, 0)
\usepackage{upquote}
\lstdefinelanguage{CSS}{
  keywords={color,background-image:,margin,padding,font,weight,display,position,top,left,right,bottom,list,style,border,size,white,space,min,width, transition:, transform:, transition-property, transition-duration, transition-timing-function},	
  sensitive=true,
  morecomment=[l]{//},
  morecomment=[s]{/*}{*/},
  morestring=[b]',
  morestring=[b]",
  alsoletter={:},
  alsodigit={-}
}
\lstdefinelanguage{JavaScript}{
  morekeywords={typeof, new, true, false, catch, function, return, null, catch, switch, var, if, in, while, do, else, case, break},
  morecomment=[s]{/*}{*/},
  morecomment=[l]//,
  morestring=[b]",
  morestring=[b]'
}
\lstdefinelanguage{HTML5}{
  language=html,
  sensitive=true,	
  alsoletter={<>=-},	
  morecomment=[s]{<!-}{-->},
  tag=[s],
    otherkeywords={
  },
  ndkeywords={
  =,
  charset=, src=, id=, width=, height=, style=, type=, rel=, href=,
  fill=, attributeName=, begin=, dur=, from=, to=, poster=, controls=, x=, y=, repeatCount=, xlink:href=,
  margin:, padding:, background-image:, border:, top:, left:, position:, width:, height:,
  transform:, -moz-transform:, -webkit-transform:,
  animation:, -webkit-animation:,
  transition:,  transition-duration:, transition-property:, transition-timing-function:,
  }
}
\usepackage{enumitem}

\usepackage[T1]{fontenc}
\usepackage{lmodern}

\begin{document}
\vspace*{0.2in}

\begin{flushleft}
{\Large
\textbf\newline{Reproducible data citations for computational research}
}
\newline
\\
Christian Schulz\textsuperscript{1*}
\\
\bigskip
\textbf{1} Computational Social Science, ETH Zurich, Zurich, Switzerland
\\
\bigskip

* cschulz@ethz.ch

\end{flushleft}
\section*{Abstract}
The general purpose of a scientific publication is the exchange and spread of knowledge. A publication usually reports a scientific result and tries to convince the reader that it is valid. With an ever-growing number of papers relying on computational methods that make use of large quantities of data and sophisticated statistical modeling techniques, a textual description of the result is often not enough for a publication to be transparent and reproducible. While there are efforts to encourage sharing of code and data, we currently lack conventions for linking data sources to a computational result that is stated in the main publication text or used to generate a figure or table.

 Thus, here I propose a data citation format that allows for an automatic reproduction of all computations. A data citation consists of a descriptor that refers to the functional program code and the input that generated the result. The input itself may be a set of other data citations, such that all data transformations, from the original data sources to the final result, are transparently expressed by a directed graph. Functions can be implemented in a variety of programming languages since data sources are expected to be stored in open and standardized text-based file formats. A publication is then an online file repository consisting of a Hypertext Markup Language (HTML) document and additional data and code source files, together with a summarization of all data sources, similar to a list of references in a bibliography. 
\section*{Introduction}

The amount of knowledge scientists produce every year is steadily increasing, as are the obstacles to clear communication of this work. The Web of Science citation index \cite{webofscience} contains almost as many publication entries for the first 15 years of the 21st century as for the whole 20th century. This is no surprise when scientists need to show their ability to consistently publish results, preferably in high-impact journals, to advance their careers. The transparency and interpretability of results are at the same time decaying due to many factors. Positive, significant results wrapped in a clear story are favored, possibly at the expense of a complete description. Furthermore, with scientific progress, the complexity of new findings naturally rises. Finally, precisely for quantitative research, the digital transformation creates new potential and challenges. With an unprecedented availability and scale of data \cite{lazer2009life}, scientists make use of \textit{Data Mining} and \textit{Machine learning} techniques that may result in comprehensive computational experiments that are difficult to protocol within the framework of traditional publications.

Ideally, a publication is completely transparent about all of its computational steps. It has been suggested that reproducibility should be a minimum standard in the computational sciences \cite{peng2011reproducible,sandve2013ten}. Reproducibility means that with the same data and methods, the same result as stated in the publication is achievable. Therefore, code and data should be provided along with the paper \cite{stodden2016enhancing}. This does not guarantee replicability (same method with new data), and ultimately, generalizability, but at least enables a first verification of reported results.

Here, a publication format is proposed that ensures automatic reproducibility and is especially suited for publishing complex computational findings. Any computational result stated in the main text of the paper needs to be linked to its data sources. A publication written with this condition does not need to be structured or laid out differently from what would be expected from a typical scientific article. This backward compatibility ensures that a paper can still be submitted to any journal that does not yet adopt these principles. Journals that want to appeal to a broad audience often even ask for a separate methods \& materials section to not obstruct the reading flow with technical details. The problem is that a typically brief description, together with mathematical notation, cannot completely communicate the actual comprehensive data transformations. With the wide-spread usage of statistical toolboxes, it becomes trivial to make a large number of methodological choices without a detailed discussion of their appropriateness and selection of parameters. Therefore, connecting the article with the actual code and data helps the reader to dig deeper when there is ambiguity or difficulty to understand. Furthermore, not everything that is coded needs to be explained in the main text. Programming environments increasingly allow for more concise code, which is potentially more understandable than an ambiguous verbal description.

The idea of such executable papers has been established since the beginnings of electronic publishing \cite{claerbout1992electronic}. SHARE \cite{van2011share} packages the complete programming environment into a virtual machine, which guarantees successful execution on another computer, but does not try to come up with a standardized way of linking publications with data and code. Another proposal is to accompany each computational result of a paper by a verifiable identifier that is created from a repository containing the code and data that generated the result \cite{gavish2011universal}. In contrast, our goal is to transparently document the sequence of steps necessary to generate the result. In this vein, \cite{peng2009distributed} caches intermediate results in order to avoid long execution times. However, it is limited to a single programming language. Finally, knitR \cite{xie2015dynamic} can be used to generate reports from a mix of programming and documenting code, but these are not easily transformed to article formats suited for submission to scientific journals.

The remainder of this paper is organized in three sections. First, the concept of \textit{data citations} is introduced to formulate a reproducible dependency graph that links all computations. Second, an implementation is described that supports the use of these \textit{data citations}. And lastly, the paper closes with a discussion of the feasibility of such an approach.

\section*{Concepts}

\subsection*{Model of computation}

\begin{figure}
\begin{adjustwidth}{0.0cm}{-0mm}
\includegraphics[width=12.0cm]{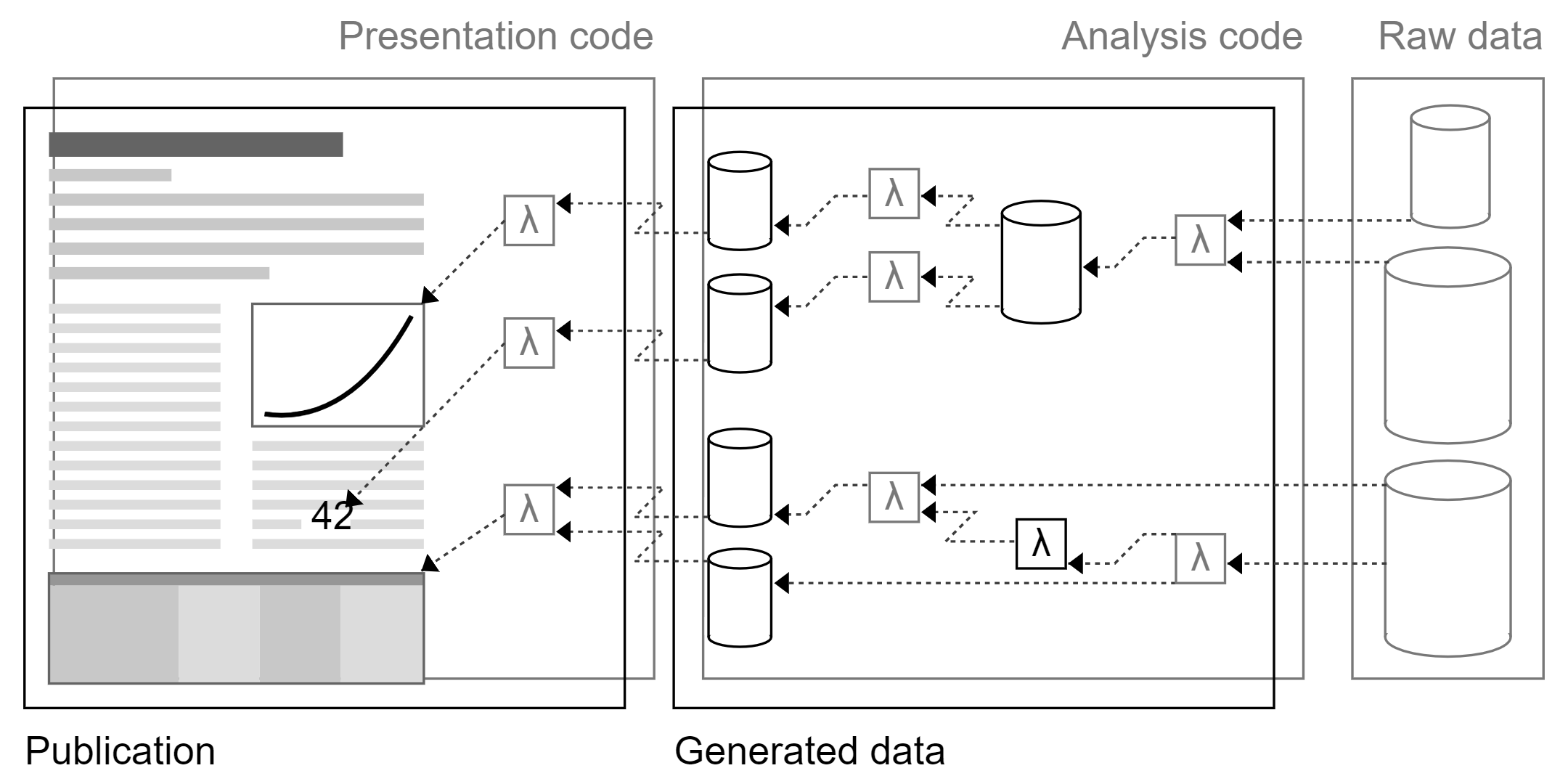}
\end{adjustwidth}
\caption[Model of computation]{\textbf{Model of computation.}  An example of a reproducible publication. Data sources are depicted as cylinders. All code is formulated as functions $\lambda$ which take data sources or parameters as an input and provide new data sources as an output. }
\label{compModel}
\end{figure}

Similar to a bibliographic citation of other published sources, a claim of a computational result should reference its origin. Such a computational citation refers to a data source that may be created from other data sources. For example, in an empirical study, we would start with measured, collected or generated data and perform an analysis that transforms them to an aggregated, numerical result which is then reported in the publication. Assuming a single computation step, it would be sufficient to provide all input and output data and a function (i.e., a computer program) that maps the input to the output. However, most papers refer to multiple computational results or explain results in a set of plots, tables and numerical statements within the main publication text. Additionally, generating a result might involve a sequence of processing steps. Fig.~\ref{compModel} outlines an example of a publication that consists of a set of computational trees with parts of the paper representing the roots and original data sources forming the leaves of the trees. Such a computational tree is a directed acyclic graph, where vertices are described as a result of a function and the input is given by the output of other results or a set of parameters. The code needs to be encapsulated in functions. On this level of abstraction, principles of functional programming are applicable independent of the actual implementation of a function, which can follow any programming paradigm and be written in most programming languages. For a reproducible paper it is required that the provided code consists of pure functions, i.e., the same input always leads to the same output (referential transparency) without side-effects (it does not modify state outside of the scope of the function). Furthermore, a function may output new functions, such that other higher-order functions may take them as input. For example, a function creates a model from training data, which is itself a function that is then applied to different data to test its predictive power. 

\subsection*{Data citations}

Besides code and data, an author of a publication also needs to provide a description that connects all pieces of information so that all computations can be automatically repeated. Each data file must be accompanied by metadata such that it can be regenerated by a referenced function and its parameters. A set of these data source descriptors can then used to build the computational tree outlined in the previous section. Data source descriptors have the following structure, resulting in a \textit{JavaScript Object Notation} (JSON) document: 

\begin{lstlisting}
{
    (<source uri> : {
        "type": <data format>,
        "func": <function uri>,
        "env": <execution environment>,
        ("code": <program code>,)?
        ("nostore": true,)?
        "params": {
            (<parameter name> : {
                "type": <data format>,
                ("parallel": true,)?
                "val": <argument value> | "uri": <source uri>
            })*
        }
    })+
}
\end{lstlisting}

\begin{itemize}[noitemsep]
\item \textit{<source uri>}: A data source URI (uniform resource identifier) is a unique path relative to the project directory and refers to the storage location of the function output. It may contain several comma-separated URIs in case of multiple outputs. A URI can be used to refer to this data source as the input of another data source generation. Only the original raw data sources do not need such a formal descriptor. Results can be retrieved at the URI once it is computed.
\item \textit{<data format>}: The format in which data is being created or interpreted when specified as an input. Valid formats are explained below.
\item \textit{<function uri>}: Identifier that refers to a function implementation provided in a specific programming language.
\item \textit{<execution environment>}: Identifier that refers to a programming environment that can execute the specified function. These environments are separately configured and provide a build automation that, for example, includes external libraries.
\item A \textit{nostore} flag, which indicates that there should be no persistent storage of this data source and it should only be computed when referred by other data sources.
\item \textit{<parameter name>}: Name used in function definition. Parameters are not ordered.
\item \textit{<argument value>}: Can be directly provided in JSON notation (which includes simple numeric or text values) or as any other data format encoded as a JSON string. Alternatively, a URI of another data source can be specified as an input. Wildcard characters may be used to refer to several URIs that can be merged to a single unordered stream or table input.
\item A \textit{parallel} flag may be specified to allow the input to be split to be processed by multiple processors, when possible.
\end{itemize}

For all data sources, there needs to be an agreement on how data is stored and can be parsed to serve as input of another function. The exact specification of possible data formats is central for building computational trees independent of programming languages, libraries, and implementations. Therefore, all original, intermediate or resulting data should be made available in non-proprietary, platform-independent data formats, both human-readable and machine-readable. Here, a list of formats is compiled that should cover most needs for scientific data:

\begin{itemize}[noitemsep]
\item JSON as defined in \cite{ecma4042013}. Can represent primitive data types as well as complex hierarchical data structures. In contrast to the comparably popular \textit{Extensible Markup Language} (XML), it is less verbose, and provides a simple syntax for lists and maps. In principle, JSON alone would be sufficient as the only data exchange format. However, for convenience and computational efficiency, further data formats are included.
\item JSONL: Sequence of lines of valid JSON values (separated by a line break). Each line can be parsed independently. Each JSON line typically represents an instance of the same data structure. It is suitable for large data sets enabling distributed computing and data stream processing.
\item CSV as defined in \cite{rfc41802005}. Contrary to the specification, a header is always required to simplify usage. Tabular data is commonly used in scientific analysis. Typically, a row represents an observation and each column a variable.
\item TXT: Any other text file format that is interpreted as a sequence of Unicode characters (encoded in UTF-8). Instead of supporting further text-based data formats such as XML, \textit{Hypertext Markup Language} (HTML), \textit{Scalable Vector Graphics} (SVG) or \textit{Resource Description Framework} (RDF) serializations directly, it is up to the user-defined function to decide for an appropriate parsing method.
\item BIN: Any binary file format. Interpreted as sequence of bytes. While we would prefer to only have human-readable text-based formats, for some purposes binary data such as \textit{Portable Network Graphics} (PNG) or \textit{Joint Photographic Experts Group} (JPEG) image files are more appropriate. Applications include image processing or a plot output, when text-based vector graphics such as SVG is not the best choice.
\item FUNC: A functional object for the use in higher-order functions. The only data type that is not required to be serializable (i.e. stored in a persistent file system).
\end{itemize}

\subsection*{Web-based publishing}

To allow for an automatic reproduction of all results, we need to re-assess the suitability of current publication format technologies. The most widely-spread document types, LaTeX and Word, usually compiled to the \textit{Portable Document Format} (PDF), cannot easily satisfy this criterion. While both provide excellent tools to produce printable journal articles, I propose that the focus should be on electronic web-based publishing instead. By relying on one of the core web technologies, HTML \cite{html2017}, with its long history of standardization and its wide range of applications, almost any kind of user interface can be represented. The simple concept of hyperlinks makes it an ideal choice for creating scientific publications that need to refer to data and code sources in a reproducible way. Writing articles directly in HTML has the advantage of adding semantic features such as the proposed \textit{data citations}. Other markup languages (such as Wiki) may be simpler to start with but could lack crucial elements of a scientific publication. HTML is undoubtedly more complicated, yet easy to learn when only utilizing a subset of features that are needed for writing articles instead of more involved web applications.

 For reproducibility, an article written in HTML is published together with code and data that are linked to the main text. It is supplied as a self-contained project folder that includes the article, code, and data with an arbitrary sub-folder structure. It can be a folder of a computer file system or published on a web server under a permanent URL. A single directory for all project content and the use of text-based file formats also simplify the use of version control systems.  

\section*{Implementation}

\begin{figure}
\begin{adjustwidth}{0.0cm}{-0mm}
\includegraphics[width=12.0cm]{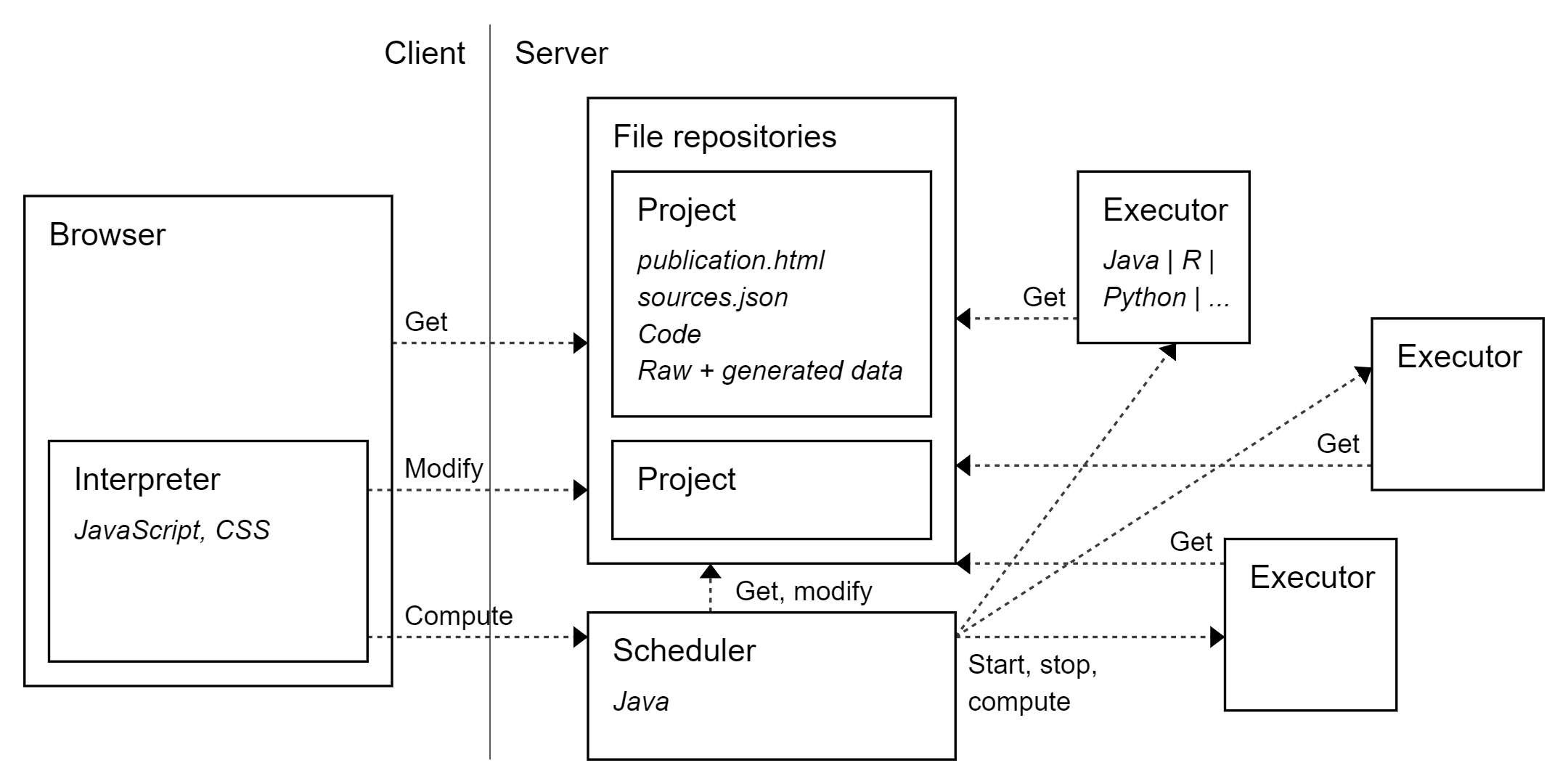}
\end{adjustwidth}
\caption[System architecture]{\textbf{System architecture.}  Articles are published as project folders that contain the publication text as HTML, all computed data and code files. These projects could be served by a standard HTTP web server. Additionally, computations as specified by the project's data source descriptors (sources.json) can be scheduled to run on potentially distributed executors. All arrows depict communication via HTTP. Computations and modifications may be subject to user authentification and only available prepublication. }
\label{systemarch}
\end{figure}

This section proposes a software design for integrating reproducible \textit{data citations} into a web-based scientific publishing system. Fig.~\ref{systemarch} outlines the system architecture.   In principle, any standard web server should be able to publish a reproducible article since in its final version it only consists of a set of static files that can be downloaded and viewed by an Internet browser. Writing publications should be as simple as possible. Therefore, the HTML document imports JavaScript libraries and \textit{Cascading Style Sheets} (CSS) to provide functionality that is commonly used in scientific writing. This includes math formulas, generation of a reference list and the separation of content from presentation. It is also possible to edit any file of a project directly in the browser and then update it server-side. Thereby, multiple authors can join an online collaboration. A layer of authentication can control pre- and post-publication viewing and modification rights.

A file named \textit{sources.json} lists all data source descriptors of a project and should be referenced by the article HTML document. Conceptually, this is similar to a \textit{BibTeX} file used for literature references. A browser can trigger a computation of a data source, either because it is missing or needs to be overwritten due to changes in code or input sources. On the server, a \textit{Scheduler} manages these computation requests and delegates them to \textit{Executors}, which perform the actual computation. Each \textit{Executor} is itself a server and is started and stopped by the \textit{Scheduler}. They are separated from the file server and \textit{Scheduler} since they are potentially implemented in a different programming language, and also need a project-specific configuration for including code, external libraries, and other parameters. All communication between different parts of the system is performed using the \textit{Hypertext Transfer Protocol} (HTTP) and according to the principles of \textit{Representational state transfer} (REST).

\subsection*{Executors}

The role of an \textit{Executor} is to perform a computation as specified by a data source descriptor. Since many programming languages are suitable for scientific computing, a scientist should be able to use any programming tool that supports best the particular task. Often, a specific statistical library that is needed is only available in a particular programming language, or acceptable performance is just feasible by using the tools provided by another programming environment. A mixed use of different programming languages within a single project is possible since functions are described solely by their input and output data formats. Each project defines a set of execution environments, which need to start a server program that is provided in different programming languages. Here, implementations for Java, Python and R are discussed. To facilitate support for additional programming languages, the required interface for a new implementation is kept as simple as possible. The primary requirement is that an HTTP server can respond to computation requests as specified by a data source descriptor. This means that there must exist a mapping of (1) all data formats to an in-memory representation (deserialization) and (2) an output back to a persistable representation (serialization). An overview is given in Table \ref{datamapping}.

In Java, a function reference is found using Java reflection. In contrast to the other dynamic languages, Java's static type system requires the input to be an instance of the parameter's type. For a conversion to and from JSON, Google's \textit{Gson} library is utilized, which provides a comprehensive mapping, even in the case of Java's generic types. JSONL maps to java.util.Stream, where each line of the input deserializes to the specified type of the stream. It is then up to the function implementation whether to load the complete data into memory, to process the stream element-wise, or not to use the stream at all. For CSV data, a new data type is provided, which allows operations on tabular data similar to built-in types found in other programming languages. In Python, JSON values can be mapped to built-in types using the standard library. JSONL data is provided by a Python generator. CSVs are directly loaded in a DataFrame of the \textit{Pandas} library. In R, most data formats can be expected to be converted to a data frame object. 

\begin{table}
\caption[Data mapping]{\textbf{Data mapping.}  Mapping of data formats to and from type representations specific to a programming language. If no library is specified, the respective standard library is used.}
\label{datamapping}
\resizebox{\textwidth}{!}{%
\begin{tabular}{ lllllll }
\hline
File format & \multicolumn{2}{c}{Java} & \multicolumn{2}{c}{Python} & \multicolumn{2}{c}{R} \\
 & type & library & type & library & type & library \\
\hline
JSON & Any & Gson & Any &  & data.frame & jsonlite \\
JSONL & Stream<T> &  & Generator &  & data.frame & jsonlite \\
CSV & Table & Koral & DataFrame & Pandas & data.frame &  \\
Other text data & Stream<String> &  & Generator &  & char vector &  \\
Binary data & byte[] &  & bytes &  & raw vector &  \\
\hline
\end{tabular}}
\end{table}

\subsection*{Scheduler}

The \textit{Scheduler} manages all computation-related tasks. The computation itself is delegated to an \textit{Executor}. When a computation of a single or a set of data sources is requested, all recursively dependent data sources need to be computed as well, in an order determined by the dependency graph. Each data source specifies an execution environment identifier under which the function can be called. A project needs to supply a command line script that takes the environment identifier as an input, starts an \textit{Executor} and returns its server URL. The \textit{Executor} could be run on the same machine as the scheduler, using a different server port, or another machine to distribute computations. While \textit{Executors} in the form of a server program are provided in a variety of programming languages, it is up to the project to configure source code inclusion or translation, library dependencies, amount of memory to be reserved or a multi-machine setup. An \textit{Executor} may be re-used by the \textit{Scheduler} for multiple computations, since functions are required to have no side-effects.

\subsection*{Interpreter}

A publication is an HTML document that loads a JavaScript library to simplify scientific writing and formatting. HTML can achieve a separation of content and styling through the use of CSS and a dynamic modification of the HTML document object model via additional scripts written in \textit{JavaScript} (JS). An author should be primarily concerned with creating the content in HTML, while more complex CSS and JS is provided, also possibly specific to the publishing journal. The following is a sample HTML publication:

\begin{lstlisting}
<!DOCTYPE html>
<html>
    <head>
        <meta charset="UTF-8">
        <script src="http://www.koral.xyz/koral.js"/>
    </head>
    <body>
        <article class="koral">
            <h2>Heading</h2>
            <h3>Subheading</h3>
            <p>
                Paragraph. 
                $x=1$.
                See <a href="#firstFig"/>.
                Given the findings of <a href="#lee2017"/>.
                The result is <span class="number" data-url="sum.json"/>.
            </p>
            <figure id="firstFig" >
                <span class="htmlpart" data-url="results/plot1.svg"/>
                <figcaption>Caption</figcaption> 
            </figure>
            <table id="mytable">
                <caption>caption</caption>
                <span class="htmlpart" data-url="result/table1.html"/>
            </table>
            <h3>References</h3>
            <div class="references" data-url="literature.bib"/>
            <h3>Sources</h3>
            <div class="sources" data-url="source.json"/>
        </article>
    </body>
</html>
\end{lstlisting}

The publication contains the following elements:

\begin{itemize}[noitemsep]
\item \textit{<h[1..6]>}, \textit{<p>} to structure the text,
\item \$...\$ for including math expressions using TeX syntax,
\item \textit{<a>} to link to bibliometric references, figures, tables, equations etc.,
\item <span data-url="..."/> to include computational results,
\item and \textit{<div class="references|sources" data-url="..."/>} for importing bibtex literature references and a list of data source descriptors.
\end{itemize}

\section*{Discussion}

\subsection*{Usability}

Writing HTML documents requires basic programming experience, which authors of papers in the computational sciences should already possess.   Even in a team of authors with mixed backgrounds, editing text passages is no more difficult than, for example, with LaTeX. However, HTML, CSS, and JavaScript can get relatively complex. Therefore, it might be necessary to agree on what features can be used for writing scientific articles. Another complication is that HTML is not a static specification; future changes of the standard might make documents no longer viewable. Besides, LaTeX's page-centric layout, is not directly available in HTML, although only relevant for printing.   On the programming side, the requirement to structure code into functions may be seen as an additional burden that is more complicated to realize than just writing a single procedural script that can access a common global state. That aside, such a high-level structure could improve the overall quality of code. It also does not require in-depth knowledge of functional programming, since within a function implementation, any programming paradigm may be used.

\subsection*{Performance}

All communication is conducted via HTTP, which may be seen as a limiting factor. However, given sufficient communication bandwidth, it is data serialization and deserialization that represent the main bottleneck. These are dependent on the utilized libraries of the respective programming language. Also, latency is not an issue, since a data source is typically a result of a long-running computation, where communication overhead is insignificant. In the case of a high number of short-running function calls within a higher-order function, they can be performed directly in the same process, given that function objects were created in the same \textit{Executor}. In the browser, loading of an HTML document, fetching all of its linked sources and formatting might take some time, depending on the complexity of the paper. Here, a solution would be to supply a pre-rendered document once the article is ready for publication.

\subsection*{Adoption}

Perhaps the greatest obstacle is adoption. Scientists may not want to publish code since with full transparency it could be easier to point out flaws in their work. For example, changing a parameter or input data could lead to a failure to arrive at the original conclusions. On the other hand, with automatic reproducibility, it should also be more straightforward for the authors to modify their experiments, and then state more clearly the limitations of their work and under which conditions the results hold. A badge system that rewards authors for being completely transparent \cite{kidwell2016badges} could be used to incentivize the practice. In some cases, publishing code or data may be perceived as losing a competitive advantage over other researchers. Additionally, some data may not be publishable due to copyright or privacy considerations. In contrast, in the presented system, both code and data publishing is not a binary decision. The representation as a computational tree allows the authors to decide at which depth level code and data are included. For example, at a certain level, data is aggregated enough that no personal information about the subjects is revealed.

 On the publisher side, such a system requires that scientific journals adapt their technical infrastructure. This is a substantial barrier. However, since the presented publication format aims to be superficially indistinguishable from established formats, an intermediate solution could be to offer a complementary publication repository that supports the proposed functionality, similar to existing pre-print servers such as \textit{arXiv}, alongside submission of manuscripts to traditional publishers.

\bibliography{literature}

\begin{thebibliography}{10}

\bibitem{webofscience}
{\relax Clarivate Analytics}. Web of Science citation index; 2017.
\newblock Available from: \url{https://webofknowledge.com}.

\bibitem{lazer2009life}
Lazer D, Pentland AS, Adamic L, Aral S, Barabasi AL, Brewer D, et~al.
\newblock Life in the network: the coming age of computational social science.
\newblock Science (New York, NY). 2009;323(5915):721.

\bibitem{peng2011reproducible}
Peng RD.
\newblock Reproducible research in computational science.
\newblock Science. 2011;334(6060):1226--1227.

\bibitem{sandve2013ten}
Sandve GK, Nekrutenko A, Taylor J, Hovig E.
\newblock Ten simple rules for reproducible computational research.
\newblock PLoS computational biology. 2013;9(10):e1003285.

\bibitem{stodden2016enhancing}
Stodden V, McNutt M, Bailey DH, Deelman E, Gil Y, Hanson B, et~al.
\newblock Enhancing reproducibility for computational methods.
\newblock Science. 2016;354(6317):1240--1241.

\bibitem{claerbout1992electronic}
Claerbout JF, Karrenbach M.
\newblock Electronic documents give reproducible research a new meaning.
\newblock In: SEG Technical Program Expanded Abstracts 1992. Society of
  Exploration Geophysicists; 1992. p. 601--604.

\bibitem{van2011share}
Van~Gorp P, Mazanek S.
\newblock SHARE: a web portal for creating and sharing executable research
  papers.
\newblock Procedia Computer Science. 2011;4:589--597.

\bibitem{gavish2011universal}
Gavish M, Donoho D.
\newblock A universal identifier for computational results.
\newblock Procedia Computer Science. 2011;4:637--647.

\bibitem{peng2009distributed}
Peng RD, Eckel SP.
\newblock Distributed reproducible research using cached computations.
\newblock Computing in Science \& Engineering. 2009;11(1):28--34.

\bibitem{xie2015dynamic}
Xie Y.
\newblock Dynamic Documents with R and knitr. vol.~29.
\newblock CRC Press; 2015.

\bibitem{ecma4042013}
{\relax Ecma International}. Standard ECMA-404: The JSON Data Interchange
  Format; 2013.
\newblock Available from:
  \url{http://www.ecma-international.org/publications/files/ECMA-ST/ECMA-404.pdf}.

\bibitem{rfc41802005}
{\relax Internet Engineering Task Force}. RFC 4180: Common Format and MIME Type
  for Comma-Separated Values (CSV) Files; 2005.
\newblock Available from: \url{https://tools.ietf.org/html/rfc4180}.

\bibitem{html2017}
{\relax World Wide Web Consortium}. HTML 5 Recommendation; 2017.
\newblock Available from: \url{https://www.w3.org/TR/html/}.

\bibitem{kidwell2016badges}
Kidwell MC, Lazarevi{\'c} LB, Baranski E, Hardwicke TE, Piechowski S,
  Falkenberg LS, et~al.
\newblock Badges to acknowledge open practices: A simple, low-cost, effective
  method for increasing transparency.
\newblock PLoS Biology. 2016;14(5):e1002456.

\end{thebibliography}

\end{document}